\documentstyle[aps,amssymb,pre,multicol,psfig]{revtex}
\begin{document}
\draft   
\author{I.V. Barashenkov\cite{igor}}
\address{Department of Physics, University of
Crete, P.O. Box 2208, 71003 Heraklion,  Greece}
\author{and E.V. Zemlyanaya\cite{elena}}
\address{Laboratory for Computing Techniques and Automation,
Joint Institute for Nuclear Research, Dubna 141980, Russia}

\title{
Existence threshold for the ac-driven damped nonlinear
Schr\"odinger solitons
}

\date{\today}
\maketitle
\begin{abstract}
It has been known for some time that  solitons of the
  externally driven, damped nonlinear Schr\"odinger equation
can only exist if the driver's strength, $h$, exceeds 
approximately $(2/ \pi) \gamma$,
where $\gamma$ is the dissipation coefficient. Although this perturbative
result was expected to be correct only to the leading order
in $\gamma$, recent  studies have demonstrated that the formula
$h_{\rm thr}= (2 /\pi) \gamma$ gives a  remarkably accurate
description of the soliton's existence threshold prompting
suggestions that it is, in fact,  {\it exact\/}.  In this note
we evaluate the next  order in the expansion
of $h_{\rm thr}(\gamma)$ showing that the actual reason for this phenomenon is
simply that the next-order coefficient is anomalously small:
$h_{\rm thr}=(2/ \pi) \gamma + 0.002 \gamma^3$.
Our approach is based on a singular perturbation
expansion of the soliton near the turning point;
it allows to evaluate $h_{\rm thr}(\gamma)$ to all orders in $\gamma$
and can be easily reformulated for other perturbed soliton
equations.
\end{abstract}

\vspace{1mm}

\begin{multicols}{2}

\section{Introduction}
 The externally driven, damped nonlinear Schr\"odinger (NLS) equation,
\begin{equation}
i \Psi_t + \Psi_{xx} + 2 |\Psi|^2 \Psi = -i \gamma \Psi -h e^{i \Omega t},
\label{NLS_0}
\end{equation}
 arises in a variety of fields 
 including plasma and condensed matter physics,
 nonlinear optics
 and superconducting electronics. 
In some of these applications (e.g. 
  in the study of the
optical soliton propagation 
in a diffractive or dispersive
ring cavity in the presence of
an input forcing beam \cite{Wabnitz}; in the description of 
 easy-axis ferromagnets in an external rotating magnetic field
 perpendicular 
to the easy axis \cite{magnetism}; in
the theory of rf-driven waves in plasma \cite{plasma})
 Eq.(\ref{NLS_0}) has a direct
interpretation. In others --- like for instance in
charge-density-wave conductors with external 
electric field \cite{CDW};
shear flows in nematic liquid crystals \cite{Lin};
 ac-driven
long Josephson junctions \cite{Josephson},
and
periodically forced Frenkel-Kontorova chains \cite{FK} ---
it occurs as an amplitude equation 
for small and slowly changing solutions of the 
externally driven, damped sine-Gordon equation:
$$
q_{tt} + \lambda q_t -q_{xx} + \sin q=
\Gamma \cos(\omega t).
$$

Without loss of generality  $\Omega$ in Eq.(\ref{NLS_0}) 
can be normalized to unity \cite{BBZ,Terrones,BS};
hence, the driver's strength $h$ and dissipation coefficient $\gamma$
are the only two essential control parameters.
Given some $h$ and $\gamma$, a fundamental question is
what  nonlinear attractors will arise
at this point of the $(\gamma, h)$-plane.
In their pioneering paper \cite{KN} Kaup and Newell 
considered Eq.(\ref{NLS_0}) on the infinite line under the vanishing
boundary conditions at infinity. By means of the Inverse Scattering-based 
perturbation theory, these authors have demonstrated
that for small $h$ and $\gamma$ Eq.(\ref{NLS_0})
exhibits two soliton solutions phase-locked to the frequency of the driver.
As $h$ is decreased for the fixed $\gamma$, the two solitons approach
each other and eventually merge in a turning point 
for $h=(2/ \pi) \gamma$ \cite{KN}. Consequently, this value plays the role
of  a threshold; no solitons exist below $h=(2/\pi)\gamma$.
Later  the same existence threshold was reobtained 
by Terrones, McLaughlin, Overman and Pearlstein  \cite{Terrones} in a regular
perturbation construction of solutions to (\ref{NLS_0})
in powers of $h$ and $\gamma$ 
(see also \cite{Spatschek}).

In ref.\cite{BS} equation (\ref{NLS_0}) was studied, numerically, in
the full range of $h$ and $\gamma$. It was found that the two
soliton solutions persist for $\gamma$ up to approximately $ 0.7$.
For each $\gamma \lesssim 0.7$ there is a turning point at some
$h=h_{\rm thr}$ at which one branch of solitons turns into another,
and which plays the role of the lower boundary of the existence
region \cite{BSA}. Amazingly, Kaup and Newell's approximate 
relation $h_{\rm thr}=(2/ \pi)\gamma$
was found to remain valid even for not very small $\gamma$.
For example, for $\gamma=0.48$ the ratio $h_{\rm thr}/\gamma$ is
different from $2/\pi$ by only one part in a thousand \cite{BS}.

A completely different approach was put forward by 
Kollmann, Capel and Bountis \cite{KCB} who regarded Eq.(\ref{NLS_0}) as the
continuous limit of a discrete NLS equation
which they studied by means of the fixed point analysis
and the Melnikov-function method. In particular, the lower 
boundary was obtained from the tangential intersection 
of the invariant manifolds of a hyperbolic fixed point.
A remarkable accuracy of Kaup and Newell's
linear law detected in \cite{BS} as well as conclusions
of their  own Melnikov-function analysis prompted the authors of \cite{KCB}
to suggest that 
the relation
$h_{\rm thr}= (2/ \pi) \gamma$ can be {\it exact\/}, 
at least for sufficiently small $\gamma$.

The aim of the present note is to demonstrate that 
 this relation
is, in fact, {\it not\/} exact, and the actual reason why it appears to be
so accurate for small $\gamma$ is simply because the coefficient of the 
next term in the expansion of $h_{\rm thr}(\gamma)$ in powers of
$\gamma$ is {\it anomalously small\/}. We do this by reconstructing the
two solitons in the vicinity of the lower boundary of their existence
domain by means of a singular (rather than regular) perturbation expansion.
This novel expansion constitutes the main technical achievement
of our work;
its scope of applicability is obviously much wider than just
the damped-driven NLS equation  (\ref{NLS_0}).

The reason why the regular expansion is not adequate near the 
solitons' existence boundary is quite simple. The point is that 
the regular expansion
is based on the assumption that solutions can be continued along
rays $h = {\frak h} \gamma$ (${\frak h}= \mbox{const}$.)
But since the boundary is  not  {\it exactly\/} a straight line $h=(2/\pi) \gamma$
(as will be shown below, it
 slowly recedes upwards from this straight line),  
the ray $h={\frak h} \gamma$ with ${\frak h}$ slightly above $2 / \pi$
will hit the boundary at some small $\gamma=\gamma_0$. Consequently,
the regular expansion fails to continue the solution beyond $\gamma_0$.
In the singular expansion, on the other hand, the solution
is continued along a curve whose shape is calculated 
simultaneously with finding perturbative corrections to the solution 
itself.
In this way the 
existence boundary can be found to any desirable accuracy.
(In this paper we restrict ourselves to terms $\sim \gamma^3$.)
This idea should remain applicable to other perturbed
soliton-bearing equations.

The outline of this paper is as follows. 
We start by  discussing the regular asymptotic expansion
as $h$ and $\gamma \to 0$ (section \ref{regular}). The procedure is similar to
the one in \cite{Terrones}; the only difference is that since   we now 
deal with solutions decaying at infinities 
($\Psi_x \to 0$) rather than periodic as in \cite{Terrones},
we will be able to find perturbative corrections in closed form.
In section \ref{singular} we explain why the perturbation series for
$\Psi$ 
breaks down as $h$ approaches  the turning point, and replace it by
a singular expansion. This  allows us to find the next terms
in the expansion of $h_{\rm thr}(\gamma)$. 
The resulting asymptotic formula is compared then with
  the threshold $h_{\rm thr}(\gamma)$ obtained in a high-precision
 numerical analysis of Eq.(\ref{stat_NLS})   
for several values of $\gamma$.
Next, after we have achieved
 an understanding of why the regular expansion fails
and how it can be cured near the existence threshold, a natural step 
is to try to develop a unified expansion which would be equally applicable
near and far from the threshold. This is done in section \ref{unified}.
Finally, some concluding remarks are made
in section \ref{conclusions} followed by a brief summary of our results.

\section{Regular perturbation expansion}
\label{regular}
By  making a substitution $\Psi(x,t)=
\psi(x,t) e^{i t}$  Eq.(\ref{NLS_0}) can be reduced  to
 an autonomous
equation
\begin{equation}
 i \psi_t + \psi_{xx} + 2 |\psi|^2 \psi - \psi = -i \gamma \psi -h.
\label{NLS}
\end{equation}
We will be interested in  time-independent
 solutions of Eq.(\ref{NLS}); these  satisfy the stationary
 equation
\begin{equation}
 \psi_{xx} + 2 |\psi|^2 \psi - \psi = -i \gamma \psi -h
\label{stat_NLS}
\end{equation}
with the boundary conditions
 $$ \psi_x(x) \rightarrow 0 \quad {\rm as} \ |x| \rightarrow \infty. $$

We start with developing a regular perturbation expansion
away from the turning point.
As the authors of  \cite{Terrones},
 we assume that we are approaching the
origin on the ($\gamma,h$)-plane along 
a straight line $h = {\frak h} \gamma$ where ${\frak h}$ is a
proportionality coefficient.
Letting 
\begin{mathletters}
\label{reg}
\begin{equation}
\psi= (u+iv) e^{-i \alpha},
\end{equation}
 where $\alpha $ is some
 constant phase that can be conveniently chosen
at a later stage, we expand
\begin{equation}
u = u_0 + \gamma u_1 +..., \quad v= v_0 + \gamma v_1+...
\label{uvexpa}
\end{equation}
\end{mathletters}and substitute into Eq.(\ref{stat_NLS}). The coefficient of
 $\gamma^0$ gives the
 unperturbed stationary NLS equation with a well-known soliton solution
\begin{eqnarray}
\left(
\begin{array}{c} u_0 \\ v_0
\end{array}
\right)
=
\left(
\begin{array}{c} \cos \theta \\ \sin \theta
\end{array}
\right)
{\rm sech} x.
\nonumber
\end{eqnarray}
Here $\theta$ is a free parameter.
Next, at the order $O(\gamma^1)$ one gets
\begin{eqnarray}
{\hat H}_0
\left(
\begin{array}{c} u_1 \\ v_1
\end{array}
\right)
= \left(
\begin{array}{c} {{\frak h}} \cos \alpha - v_0 \\
{{\frak h}} \sin  \alpha + u_0
\end{array}
\right),
\label{mat}
\end{eqnarray}
where the Hermitean operator
\begin{eqnarray}
{\hat H}_0
\equiv
\left(-\frac{\partial^2}{\partial x^2} +1\right) {\hat I}-
2 \left(
\begin{array}{lr}
 v_0^2 + 3 u_0^2
&
2u_0 v_0\\
2u_0 v_0 &
u_0^2 + 3v_0^2
\end{array}
\right)
\label{H1}
\end{eqnarray}
 and ${\hat I}$ is the $2 \times 2$ identity matrix.
In order for the equation (\ref{mat}) to be solvable
in the class of bounded functions, its right-hand side
needs to be orthogonal to the vector $(v_0, -u_0)^T$,
the eigenfunction of the operator ${\hat H}_0$
associated with the zero eigenvalue.
(This zero eigenvalue results from the U(1) phase-invariance of the unperturbed NLS equation.)
The orthogonality gives a relation between $\alpha$
and $\theta$,
\begin{equation}
\pi {{\frak h}} \sin (\theta-\alpha) =2,
\label{rela}
\end{equation}
implying that only one of the two parameters (say, $\theta$) can be chosen
freely. It does not matter what exactly  we  choose for $\theta$;
the net phase of the leading-order approximation is
$(\theta-\alpha)$ and this is fixed by Eq.(\ref{rela}).
The meaning of this relation is straightforward.
For $h=\gamma=0$,
the NLS equation has a family of soliton solutions,
$\psi=e^{i(\theta-\alpha)} {\rm sech\/}x$,
with $(\theta-\alpha)$ arbitrary. However, if we
want to continue the solution along the line $h= {\frak h} \gamma$,
the unperturbed solution that we need to start with has the phase given by 
Eq.(\ref{rela}).

 It is convenient to take $\theta= \pi/2$;
this makes the linear operator (\ref{H1})
diagonal.
(The other diagonal choice 
$\theta=0$ is also acceptable but somewhat less convenient 
in the present context.)  The constant phase $\alpha$
is then determined by
\begin{equation}
\cos \alpha = \frac{2}{\pi} \frac1 {{\frak h}}.
\label{alpha}
\end{equation}
In fact, there are two values of $\alpha$ defined by this equation,
one positive and one negative. The positive
$\alpha=\alpha_+$ corresponds to the soliton $\psi^{(+)}$
and the negative $\alpha=\alpha_-$ defines the soliton $\psi^{(-)}$.
Since the left-hand side cannot exceed 1, the right-hand side gives
the well-known formula for the lower boundary of the domain
of existence of the two solitons:
${{\frak h}} \ge {{\frak h}}_{\rm thr}= 2/ \pi$
\cite{KN,Terrones,Spatschek}.
(In the next section we will obtain a more precise formula for this threshold.)

Now for $\theta=\pi/2$ the equations (\ref{mat})  become
\begin{eqnarray}
L_0 u_1(x)
=
 {{\frak h}} \cos \alpha  - v_0(x); \label{m1}
\\
L_1 v_1(x)=  {{\frak h}} \sin \alpha,    \label{m2}
\end{eqnarray}
where $v_0(x)={\rm sech\/} x$
and $L_0$ and $L_1$ are the standard Schr\"odinger operators with familiar
spectral properties:
\begin{eqnarray}
L_0= - \partial^2/\partial x^2 + 1-2{\rm sech\/}^2 x; \label{L0}\\
L_1=-\partial^2/\partial x^2+1 - 6 {\rm sech\/}^2x.
\label{L1}
\end{eqnarray}
The operator $L_1$ is invertible
on even functions; in particular,
\begin{equation}
L_1^{-1} {\rm sech} \, x =  \frac12 (x \,
{\rm tanh} \, x -1) {\rm sech\/}x,
\label{prop0}
\end{equation}
\begin{equation}
L_1^{-1} {\rm sech}^3 x = - \frac14 {\rm sech} x
\label{prop1}
\end{equation}
and
\begin{equation}
L_1^{-1} 1 = 1- 2{\rm sech}^2 x.
\label{prop2}
\end{equation}
 Hence Eq.(\ref{m2}) is readily solved:
\begin{equation}
v_1= {{\frak h}} \sin \alpha  (1 - 2 {\rm sech}^2 x).
\label{v_1}
\end{equation}
The condition (\ref{alpha}) being in place, Eq.(\ref{m1}) is solved
as well:
\begin{mathletters}
\label{u1}
\begin{equation}
u_1= {\cal U}_1(x) + A {\rm sech\/}x,
 \label{ua}
\end{equation}
where
\begin{eqnarray}
{\cal U}_1(x)= \frac{2}{\pi} +
 \frac12 {\rm tanh }x \, {\rm sinh} x
+ \frac{1}{\pi} \times
 \{
j(x) \,   {\rm sech} x   - \nonumber \\
- ( x \, {\rm sech} x + {\rm sinh} x ) \,
{\rm arcsin} ({\rm tanh} x) -1 \},
\label{a}  \\
j(x) \equiv \int^x_0 \xi \, {\rm sech} \xi \, d \xi,
\nonumber
\end{eqnarray}
\end{mathletters}and 
$A$ is an arbitrary constant which is to be fixed at higher orders
of the expansion.
Hence we proceed to $O(\gamma^2)$ to find
\begin{eqnarray}
L_0 u_2
=
(4 v_0 u_1-1) v_1,
\label{m3}
\\
L_1 v_2= u_1+ 2 v_0 (u_1^2 + 3 v_1^2).   \label{m4}
\end{eqnarray}
Equation (\ref{m3}) is solvable if its right-hand side is orthogonal
to ${\rm sech }x$. Substituting from (\ref{v_1})-(\ref{ua}),
this condition fixes the constant $A$:
\begin{equation}
A=A^{(0)} \equiv \frac4 {\pi} \int  {\cal U}_1(x) \,
 {\rm sech}^2 x (1 -2 {\rm sech}^2 x) dx,
\label{At}
\end{equation}
where we have used
 Eqs.(\ref{prop1}-\ref{prop2}).
(Here we have written $A^{(0)}$ for $A$ so as to emphasize that 
this is now a fixed number; this number will reappear in 
the singular expansion below.)
Eq.(\ref{m3}) is now solved in the form
\begin{equation}
u_2= {\cal U}_2(x) + B {\rm sech} x.
\label{m7}
\end{equation}
The constant $B$ is to be fixed 
at the
$\gamma^3$-level, where we obtain the
equation
\begin{equation}
L_0 u_3
=
2 \{ u_1 (v_1^2 + u_1^2) +2 v_0 ( u_2 v_1 +u_1 v_2)  \}- v_2.
\label{m5}
\end{equation}
The solvability condition for eq.(\ref{m5})  gives us $B$:
\begin{eqnarray}
B=
(\pi {{\frak h}} \sin \alpha)^{-1} 
\times \nonumber \\
  \int
v_0   \{ 2u_1(u_1^2+v_1^2)
+
4 v_0 (u_1 v_2 + {\cal U}_2 v_1) -v_2
 \} \, dx.
 \label{B}
\end{eqnarray}

So far our treatment followed the lines of Terrones {\it et al}
\cite{Terrones};
the only difference is that our $v_0, u_1, v_1,...$ are given by
explicit formulas. Using (\ref{a}) in  (\ref{At}) and
integrating numerically, we identify the constant $A^{(0)}$
which completes the determination of the first-order corrections:
$A^{(0)}= -2.4378 \times 10^{-1}$. 

Let us now send
$h \to (2 / \pi) \gamma$.
The formula (\ref{u1}) for $u_1(x)$ is not affected
 and
the  expression (\ref{At}) for $A$ remains valid as well.
Therefore, the solvability of Eq.(\ref{m3}) is ensured and
$u_2$ can be written in the form (\ref{m7}).
The constant $B$ is expected to be identifiable from Eq.(\ref{B}).
However,
for ${{\frak h}} \to 2 / \pi$ we have
 $\sin \alpha \to 0$ and so
this formula gives $B = \infty$ unless
\begin{equation}
\int v_0
( 2 u_1^3+ 4 v_0 u_1 v_2  - v_2 )
dx=0.
\label{condi}
\end{equation}
(Here we have used that $v_1 \to 0$ as $\sin \alpha \to 0$.)
In general the condition (\ref{condi}) is {\it not\/} in place,
and therefore the regular expansion blows up.

\section{Singular perturbation expansion at the turning point}
\label{singular}

The reason for the breaking down of the expansion is that
it was implicitly assumed  in Eq.(\ref{uvexpa}) that $v_1 =O(1)$ whereas in the
actual fact, in the
limit ${{\frak h}} \to 2 / \pi$ we have $v_1 \to 0$. Let us now
explicitly take this fact into account by  writing
\begin{mathletters}
\label{sing}
\begin{equation}
u =  \gamma u_1 + \gamma^2 u_2..., \quad v=
v_0  + \gamma^2 v_2+...,
\label{cor}
\end{equation}
where $v_0={\rm sech} x$. We also expand ${\frak h}$:
 \begin{equation}
{{\frak h}}=  {{\frak h}}_0 + {{\frak h}}_1 \gamma + {{\frak h}}_2 \gamma^2+....,
\quad {{\frak h}}_0= \frac2{\pi}.
\label{hexp}
\end{equation}
\end{mathletters}(Thus 
we have  fixed $\theta= \pi/2$ and  $\alpha=0$ straight away.)
Substituting into (\ref{stat_NLS}), the first order in $\gamma$
yields eq.(\ref{m1}) where one   should only replace 
${\frak h} \cos \alpha \to {\frak h}_0$.
Its solution is given by the same
eq.(\ref{u1}) as before, with $A$  an undetermined constant.
At the order $\gamma^2$ we obtain
\[
L_0 u_2 ={{\frak h}}_1,
\]
and hence ${{\frak h}}_1=0$ and $u_2=B \, {\rm sech\/} x$.
The equation for $v_2$ is now
\begin{equation}
L_1 v_2 =  u_1 + 2 v_0 u_1^2
\label{v_2}
\end{equation}
[cf. Eq.(\ref{m4})]; this is always solvable. 
Finally, the $\gamma^3$-level yields 
\[
L_0 u_3= 2( u_1^3+ 2 v_0 u_1 v_2 ) -v_2 + {{\frak h}}_2
\]
[cf. (\ref{m5})],
whose solvability condition is given by
\begin{equation}
\int  v_0  \,(  2 u_1^3 +
4 v_0 u_1 v_2 
 -v_2
+ {{\frak h}}_2 ) \, dx=0.
\label{condi2}
\end{equation}

It turns out  that it is only this equation (\ref{condi2})
that fixes the constant $A$ in Eq.(\ref{ua}).
 Indeed, substituting $u_1$ from (\ref{ua}) and $v_2$ from  (\ref{v_2}),
Eq.(\ref{condi2}) reduces to a quadratic equation 
\begin{equation}
A^2 - 2 { P}A + { Q} - \pi {{\frak h}}_2=0,
\label{quadra}
\end{equation}
where after some algebra the coefficients are found to
be
\begin{equation}
{ P}= -2 {{\frak h}}_0^2  + \frac{{{\frak h}}_0}{2} \int \{ {\frak h}_0-
{\cal U}_1(x)\} dx
\label{p}
\end{equation}
and
\begin{eqnarray}
{ Q}= \int \{ {\cal U}_1^2 -{\cal U}_1
 (1+ 2 {\cal U}_1 {\rm sech\/}x) 
 \times \nonumber \\
  \times
 [ {\frak h}_0 (1 - 2 {\rm sech\/}^2x) 
 + \, {\rm sech\/}x \, ( 1- x  \, {\rm tanh\/}x)] \} dx.
   \label{q}
\end{eqnarray}
In the derivation of (\ref{p}-\ref{q}) we used Eq.(\ref{prop0}) and the 
identity
\begin{equation}
4 L_1^{-1} (v_0^2 \, {\cal U}_1) = L_1^{-1}( {\frak h}_0-v_0) -{\cal U}_1;
\label{identity}
\end{equation}
this is a straightforward consequence of Eqs.(\ref{alpha}),(\ref{m1})
and the fact that the Schr\"odinger operators 
(\ref{L0})-(\ref{L1}) differ by $4  \, {\rm sech\/}^2x$:
$$
L_0= L_1+ 4 v_0^2(x).
$$
Since there is a cubic term in  $u_1$  in Eq.(\ref{condi2}), 
one could expect 
the  resulting equation for $A$ to be cubic; however the coefficient in
front of $A^3$ is easily shown  to vanish. Another observation is
that the coefficient ${ P}$  coincides with the constant 
$A^{(0)}$ [Eq.(\ref{At})] obtained in the regular expansion. To see that, 
one only needs to use the identity (\ref{identity}) once again.

The roots of Eq.(\ref{quadra}) are given by
\begin{equation}
A^{(\pm)}= A^{(0)} \pm \sqrt{ \pi \left({\frak h}_2-{\frak h}_2^{(0)} \right)},
\label{Apm}
\end{equation}
where
\begin{equation}
{\frak h}_2^{(0)} \equiv \frac1{\pi} ({ Q}-{ P}^2).
\label{h20}
\end{equation}
Hence, if ${\frak h}_2 > {\frak h}_2^{(0)}$, Eq.(\ref{stat_NLS}) has 
 two solutions
$\psi^{(\pm)}$ which are only different  in the coefficients $A^{(\pm)}$. If
${\frak h}_2 < {\frak h}_2^{(0)}$, there are no solutions at all.
The value ${\frak h}_2={\frak h}_2^{(0)}$ is therefore the turning point.
Doing numerically the integral in (\ref{q}) we find
 ${ Q}= 6.4665 \times 10^{-2}$. 
Recalling that ${ P}$ coincides with  
Eq.(\ref{At}), ${ P}=A^{(0)}=-2.4378 \times 10^{-1}$,
Eq.(\ref{h20}) gives   
 ${{\frak h}}_2^{(0)}=1.667 \times 10^{-3}$.
Finally, the coefficient $A$ corresponding to the turning point 
 coincides with the  ``far from the turning point" 
 value, Eq.(\ref{At}): $A=A^{(0)}$.

It is worth noting here  that  if ${\frak h}_2=0$, Eq.(\ref{condi2}) is formally coincident with 
Eq.(\ref{condi}). This does not mean, however, that soliton solutions 
exist
for ${\frak h}_2=0$, and that these solutions can be found
by regular expansions (\ref{reg}). The difference 
 is that in Eq.(\ref{condi}) 
the function $u_1(x)$ has the coefficient $A$
which has already been fixed by Eq.(\ref{At}),
whereas Eq.(\ref{condi2}) is an equation for {\it unknown\/} $A$.

Next, how close to $h_{\rm thr}$ does the regular expansion stop working
and has to be replaced by the singular one? 
For ${\frak h}= 2/\pi + {\frak h}_2 \gamma^2$ Eq.(\ref{alpha}) 
produces
\begin{equation}
\alpha_{\pm} = \pm \sqrt{\pi {\frak h}_2} \gamma + O(\gamma^3)
\end{equation}
and so the regular expansion for $\psi^{(\pm)}$ reads
\begin{eqnarray}
\psi^{(\pm)} = i \, {\rm sech\/}x + \gamma {\cal U}_1(x)
+  \nonumber\\
+ \gamma \left(A^{(0)} \pm \sqrt{\pi {\frak h}_2} \right) \, {\rm sech\/}x + O(\gamma^2)
\label{wrong} 
\end{eqnarray}
with ${\cal U}_1$ as in (\ref{a}), whereas the correct, singular
expansion is
 \begin{equation}
\psi^{(\pm)} = i  \,{\rm sech\/}x + \gamma {\cal U}_1(x)
+ \gamma A^{(\pm)} {\rm sech\/}x + O(\gamma^2),
\label{right}
\end{equation}
where $A^{(\pm)}$ are given by Eq.(\ref{Apm}).
 Comparing (\ref{wrong}) to (\ref{right}), one 
 concludes that  the difference 
between the regular and singular 
expansions is negligible provided
${\frak h}_2 \gg {\frak h}_2^{(0)} \approx 2 \times 10^{-3}$. 
On the contrary, for ${\frak h}_2 \sim  10^{-3}$
 the difference
cannot be ignored. 

We close this section by comparing our asymptotic approximation
for $h_{\rm thr}(\gamma)$  with  results of the high-accuracy
 numerical solution
of Eq.(\ref{stat_NLS}). Here we employed 
a fourth-order Newtonian algorithm, with the stepsize 
$\Delta x=0.005$ and residual $5 \times 10^{-11}$.
 Eq.(\ref{stat_NLS}) was solved 
on a half-interval $(0,40)$. We would 
decrease  $h$  in increments of $\Delta h= 1 \times 10^{-10}$
until iterations
have ceased to converge; after that
we would repeat the computations on a condensed grid
and extended interval. As a result we were able to obtain
  the threshold $h_{\rm thr}$ 
 accurate to ten digits after the decimal point. 
 
 Fig.1 shows the difference between the numerically found values of
 $h_{\rm thr}$ and (a) Kaup and Newell's linear law $h_{\rm thr}
 = (2/ \pi) \gamma$, and (b) the refined
 asymptotic  expansion $h_{\rm thr}
 = (2/ \pi)\gamma + {\frak h}_2^{(0)} \gamma^3$ with 
 ${\frak h}_2^{(0)}=1.667 \times 10^{-3}$.
 The latter expansion is indeed seen to provide a more accurate 
 approximation 
 for the threshold. 
 \begin{figure}
\begin{center}
\psfig{file=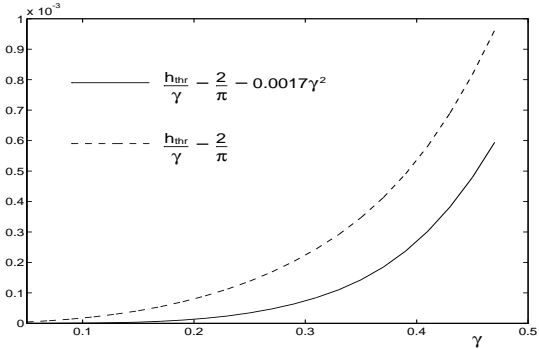,height=5cm,width=1.\linewidth}
\end{center}
\vspace{-1mm}
\caption{
The comparison of the numerically obtained threshold
values and their analytic approximations. Plotted is the 
difference $\left( h_{{\rm thr\/}(numerical)} -h_{{\rm thr\/}(analytic)}
 \right)
/\gamma$, where $h_{{\rm thr\/}(analytic)}$ is given by
 $(2/ \pi) \gamma$ (dashed curve)
and  $(2/\pi) \gamma + 0.0017 \gamma^3$ (solid line). 
}
\end{figure}

\section{A unified viewpoint}
\label{unified}

The singular expansion of the previous section was designed as a
   continuation and generalization of the regular
expansion and presented in the form that 
 allows for a straightforward comparison with  
     the latter.  
 A natural  step now is to try to  develop  
 a unified approach which would be valid both near and far from the 
 turning point. Such a unified formalism could provide an additional insight
 into the structure of solutions and have some technical advantages.
   
The unification is achieved 
if one notices  that the role of the coefficients $A,B,...$ is
simply  to renormalize
the constant phase of the leading-order approximation $\psi_0
=i {\rm sech\/}x$. 
Consequently,  instead of adding homogeneous solutions $A {\rm sech\/} x$,
$B {\rm sech\/}x,...$ at each order of $\gamma$ we can expand $\alpha$:
\begin{equation}
\alpha = \alpha_0 +  \alpha_1 \gamma + \alpha_2 \gamma^2 + ... .
\label{exp_alpha}
\end{equation} 
(This observation belongs to D. E. Pelinovsky.)
Multiplying Eq.(\ref{stat_NLS}) through by
$e^{i \alpha}$:
\begin{equation}
\psi_{xx} + 2 |\psi|^2 \psi - \psi =- i \gamma \psi - h e^{i \alpha},
\label{psalpha}
\end{equation}
and 
substituting Eqs.(\ref{hexp}), 
(\ref{exp_alpha}) and (\ref{uvexpa}) with $u_0=0$ and $v_0= {\rm sech\/}x$, 
 we get,
at the order $\gamma$: 
\begin{eqnarray}
{\frak h}_0= \frac{2}{\pi} \frac1{\cos{\alpha_0}};
\label{halpha}\\
u_1={\cal U}_1, \quad  v_1= {\frak h}_0 \sin \alpha_0 L_1^{-1} 1,
\label{new_v1}
\end{eqnarray}
 with ${\cal U}_1$  as in Eq.(\ref{a}).
For any $\alpha_0$ Eq.(\ref{halpha}) gives the  initial slope of the 
curve $h(\gamma)$ along which we want to continue our solution.
The lowest $h$ arises for $\alpha_0=0$; hence in order to
obtain the threshold one has to set  ${\frak h}_0=2/ \pi$.
Next, the order $\gamma^2$ yields
\begin{equation}
{\frak h}_1 = {\frak h}_0 \tan \alpha_0
\left\{ \alpha_1 + \frac{1}{\pi} \int v_0(1- 4 v_0 u_1) L_1^{-1} 1 dx
\right\}
\label{hal1}
\end{equation}
and 
\begin{equation}
v_2= L_1^{-1} (u_1+2 v_0 u_1^2 +  {\frak h}_0 \alpha_1).
\label{new_v2}
\end{equation} 
 Eq.(\ref{hal1}) relates ${\frak h}_1$ and 
$\alpha_1$.  One of these  (say, ${\frak h}_1$)
can be chosen freely; then Eq.(\ref{hal1}) fixes the other.
This simply means that given  a nonzero $\alpha_0$, the solution 
exists for an arbitrary ${\frak h}_1$
(i.e. in an arbitrary neighbourhood of the ray $h={\frak h}_0 \gamma$.) 
  The case $\alpha_0=0$ is exceptional; 
  in this case  the ray  is very close to the threshold and
  so we have to set ${\frak h}_1=0$ while $\alpha_1$ remains 
undetermined. (This is the case where we would have to invoke
the singular expansion before.)
 The coefficient $\alpha_1$ can only be fixed at the $\gamma^3$-level,
where we obtain 
\begin{eqnarray*}
\frac{1}{\pi}
\int v_0 \left\{ 2[ u_1(u_1^2  + v_1^2) + 2 v_0 (u_1 v_2 + u_2 v_1)]
-v_2 \right\} dx =\\
= 
{\frak h}_1 \alpha_1 \sin \alpha_0 - {\frak h}_2 \cos \alpha_0 +
{\frak h}_0 \left(\alpha_2 \sin \alpha_0 
+ \frac{\alpha_1^2}2 \cos \alpha_0 \right).
\end{eqnarray*}
Letting $\alpha_0=0$ this becomes exactly the quadratic equation (\ref{quadra})
(where we only need to replace $A \to \alpha_1$) and hence we 
have reproduced our previous threshold.

Note that in the new formalism the corrections in the phase 
of $\psi_0$ are uncoupled from the rest of the expansion.
The advantage of this is that  the correction $\alpha_1$ ($=A$ in the
previous notation) arises at the order $\gamma^2$ 
and not at $O(\gamma)$ as  before;
the correction $\alpha_2$ (previously known as $B$) arises at $O(\gamma^3)$
and not at $O(\gamma^2)$, and so forth. 
As a result, $u_1$ does not contain the unknown $\alpha_1$ ($A$);
the correction $v_2$ depends on $\alpha_1$ only linearly (not quadratically
as before), and so the resulting equation for $\alpha_1$ is
manifestly quadratic.
(Recall that Eq.(\ref{quadra}) arises initially as a {\it cubic\/}
equation and only then the coefficient of the term $A^3$ is calculated 
to be zero.)

\section{Concluding remarks and conclusions}
\label{conclusions}

\noindent
{\bf 1.} In the undamped case ($\gamma=0$) for any $h \in (0, \sqrt{2/27})$
Eq.(\ref{stat_NLS})
 has two  explicit  solutions \cite{BBZ}:
\begin{equation}
\psi^{(\pm)}(x) = \psi_{\infty} \left\{
1+ \frac{2 \sinh^2 \beta}{1 \pm  \cosh ({\cal A}x) \cosh \beta}
\right\},
\label{g=0}
\end{equation}
 \begin{figure}
\begin{center}
\psfig{file=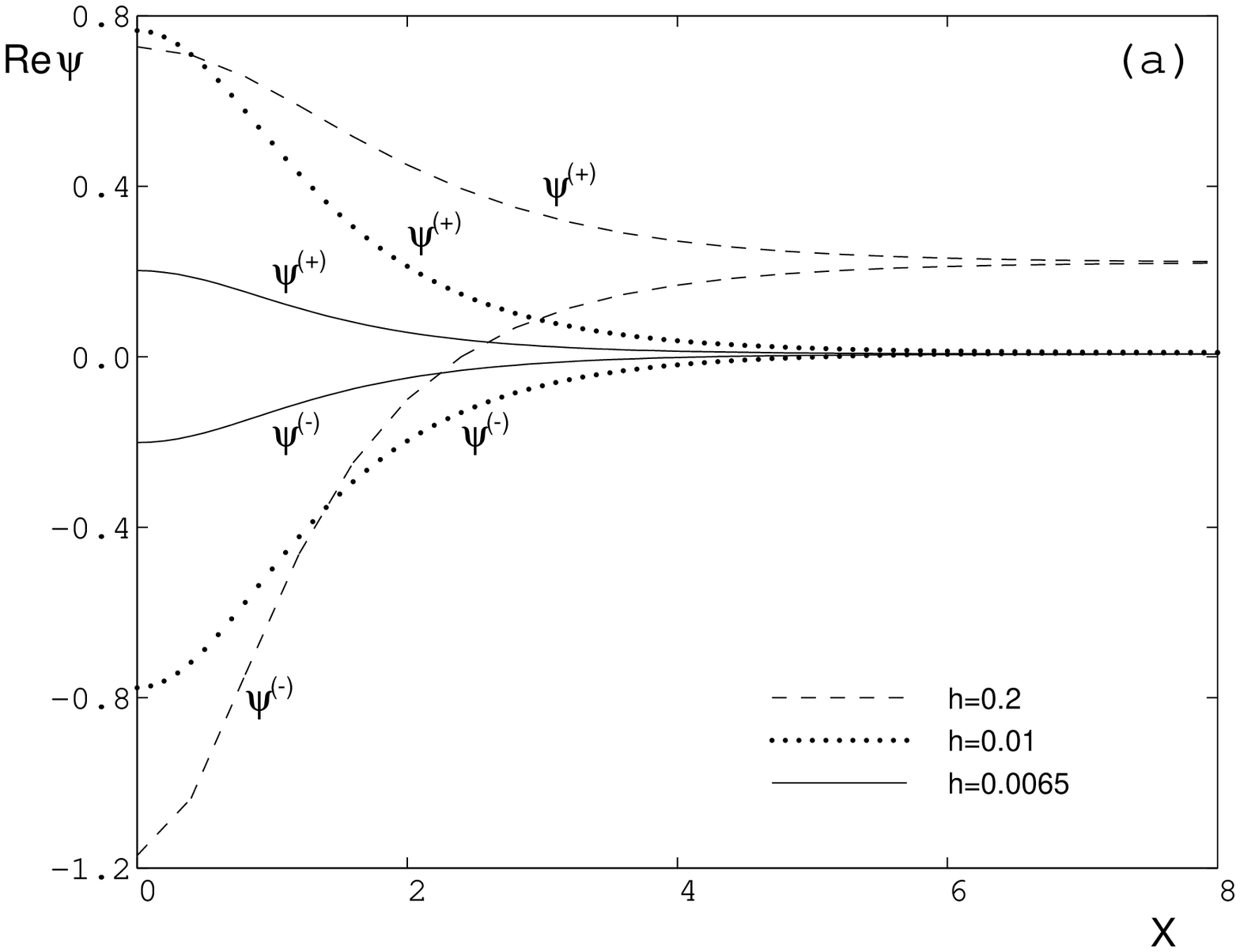,height=5cm,width=1.\linewidth}
\psfig{file=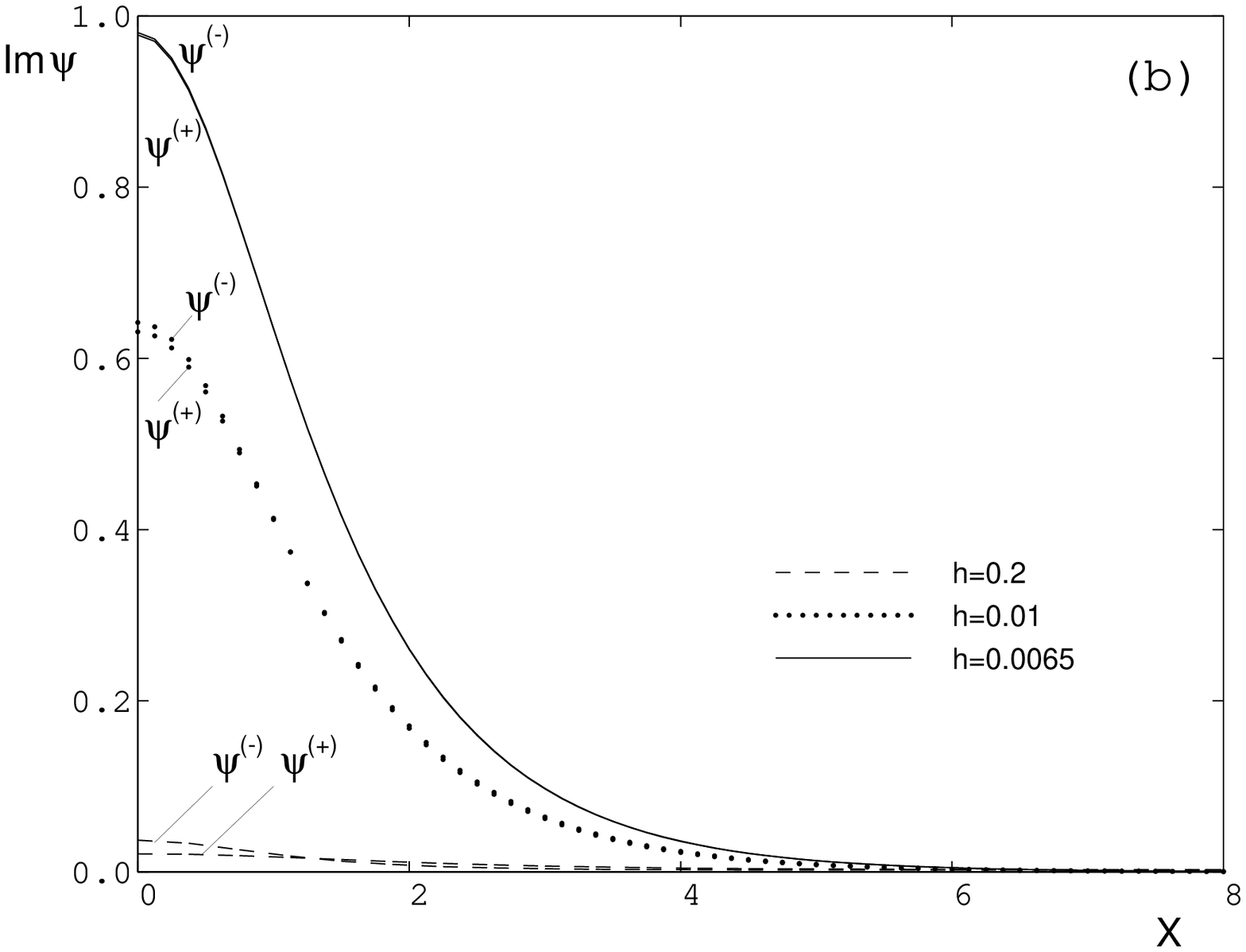,height=5cm,width=1.\linewidth}
\psfig{file=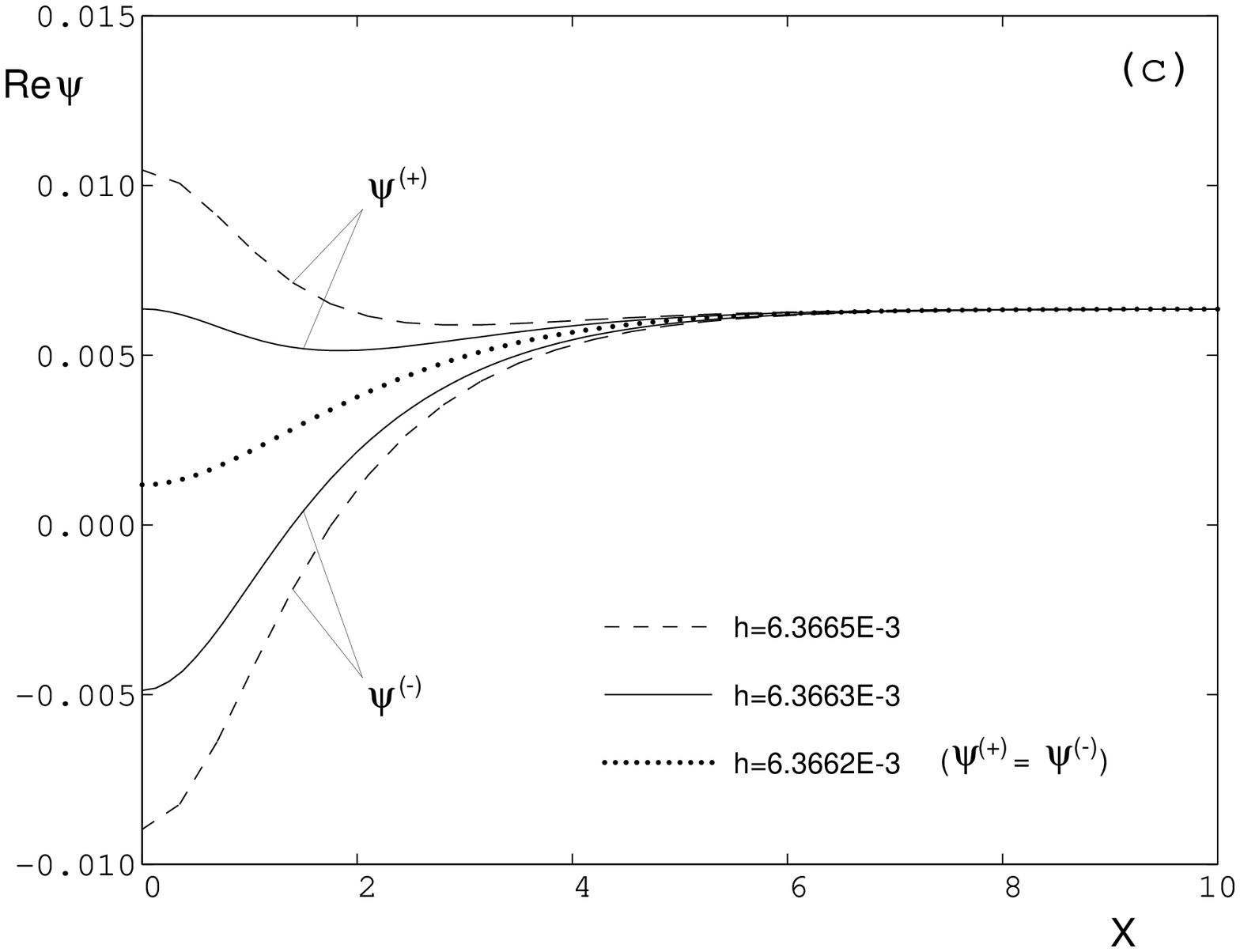,height=5cm,width=1.\linewidth}\vspace{-1mm}
\psfig{file=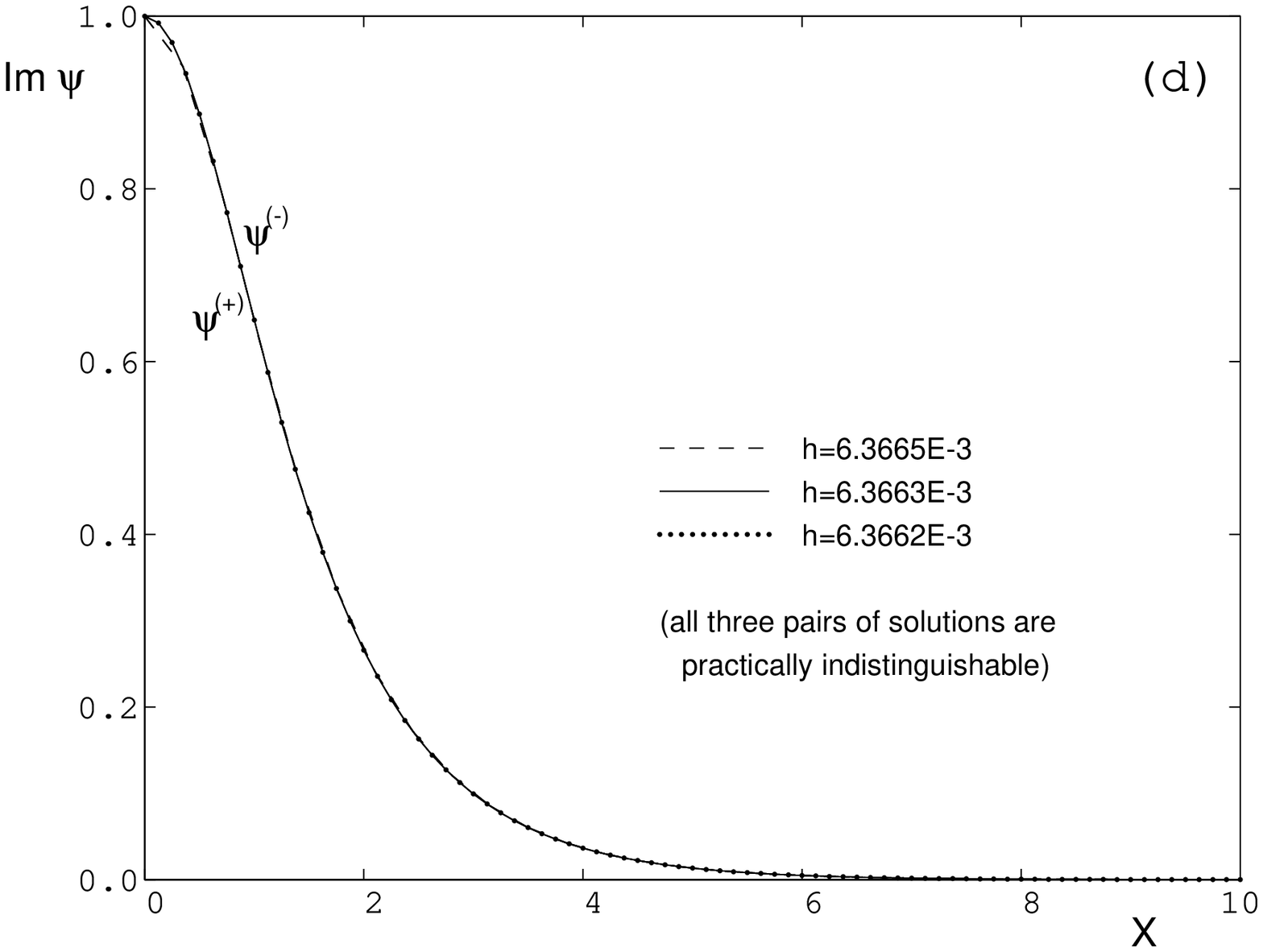,height=5cm,width=1.\linewidth}
\end{center}
\caption{
Soliton transformation for small $\gamma$
 ($\gamma=0.01$). (a,b): For $h$ far above  the turning point 
$h_{\rm thr}= 6.3661997 \times 10^{-3}$
 the imaginary parts are close to zero and the two solitons
 are well approximated by the two undamped solitons (\ref{g=0})
 (dashed lines). 
  As $h$ approaches the turning point, the real parts decrease and converge
  while imaginary parts grow (dotted then solid lines.)
  (c,d): In the immediate  vicinity of $h_{\rm thr}$ 
  the hump in the $\psi^{(+)}$ profile rapidly transforms into a dip
  and the two solitons collapse into one 
  (dashed then solid  then dotted line.)
  All functions being even, we only show them for positive $x$.
}
\end{figure}
 
where 
$$ \psi_{\infty} = \left\{ 2(1+ 2 \cosh^2 \beta) \right\}^{-1/2} $$
is the asymptotic value of $\psi^{(\pm)}(x)$ as $|x|\to \infty$; the parameter
 $\beta$ is defined by inverting the relation
$$
h= \frac{ \sqrt{2} \cosh^2 \beta}{(1+ 2 \cosh^2 \beta)^{3/2}},
$$
and ${\cal A}$ is given by
$$
{\cal A}= 2 \psi_{\infty} \sinh \beta = \frac{ \sqrt{2} \sinh \beta}
{\sqrt{1 + 2 \cosh^2 \beta}}.
$$

For the given value of $h$ the
$\psi^{(+)}$ branch merges with $\psi^{(-)}$
at some  $\gamma_{\rm thr}$ [defined, approximately, by
$h=(2/\pi) \gamma_{\rm thr} + {\frak h}_2^{(0)} \gamma_{\rm thr}^3$].
If $h$ is small then this $\gamma_{\rm thr}$ is also small,
so that the point of the merger is very close to the $h$-axis
and hence intuitively one could expect solutions at this point to be
close to the undamped solitons (\ref{g=0}). 
However, it is not quite obvious how this proximity can be reconciled 
with the fact that 
the two real solutions (\ref{g=0})
 are rather far from each other.
 
A related question concerns the shape of the $\psi^{(+)}$ and $\psi^{(-)}$
solitons. In the undamped case [Eq.(\ref{g=0})] the real function
$\psi^{(+)}$ has  a hump and $\psi^{(-)}$  has a dip.
The numerical analysis shows that the hump
respectively the dip persist in real parts of $\psi^{(+)}$
respectively $\psi^{(-)}$ 
solitons  for $\gamma  \ne 0$ \cite{BS}. Again, it is intuitively 
not quite obvious 
how  the hump can transform into dip as the two branches merge.
Shall this transformation proceed via the flat state?

The behaviour of solutions for small $h$ and $\gamma$
(in particular, near the turning point)
 can be clarified by invoking the asymptotic expansions
(\ref{reg}) and (\ref{sing}). For the sake of
illustration, we  have also computed the two solitons numerically
for a fixed small $\gamma$ ($\gamma=0.01$) and varying $h$; results are 
shown in Fig.2. In agreement with Eq.(\ref{reg}),
for not very small $h$ ($h \gtrsim 0.1$) the imaginary parts
of the solitons are seen to be almost zero  [dashed lines in Fig.2(b)] 
while the real parts  change slowly
with the variation of $h$. For these $h$ the pair of real 
solutions 
(\ref{g=0}) does indeed provide a good approximation for the corresponding 
$\psi^{(\pm)}$ with $\gamma=0.01$.
However, as $h$ goes down, the real parts start changing
(decreasing in magnitude) more vigorously while the  imaginary parts
are not small any longer; consequently, the approximation deteriorates. 
Near the turning point
the real parts of both solitons become 
much smaller than their imaginary parts [see Fig. 2(c,d)]. 
Nevertheless,
for $h$ not {\it very\/} close to the turning point
(more specifically, for $h \ge 6.3665 \times 10^{-3}$),
the real part of $\psi^{(+)}$ still 
 has a hump and real part of $\psi^{(-)}$ 
still has a dip [Fig. 2(a)]. This justifies our usage of the notations
$\psi^{(+)}$ and $\psi^{(-)}$ for $\gamma \neq 0$. Finally, in a
very near vicinity of 
 the existence threshold $h_{\rm thr}=6.3662 \times 10^{-3}$,
the hump of the real part quickly transforms into the dip [Fig.2(c)].

 This 
   rapid change 
 near the turning point
  can be easily
understood in terms of the singular
  expansion (\ref{sing}). 
When 
${\frak h}_2$ is close to
${\frak h}_2^{(0)}$, we can define a small $\epsilon$ by writing 
\begin{equation}
\pi {\frak h}_2= \pi {\frak h}_2^{(0)} + \epsilon.
\end{equation}
The corresponding $A$'s in Eq.(\ref{ua}) are then given
by 
\begin{equation}
A^{(\pm)}= A^{(0)} \pm \sqrt{\epsilon}.
\label{aeps}
\end{equation}
Recalling that to the leading order in $\gamma$ it is only this coefficient
$A$  that determines the dependence of solutions on $h$,
Eq.(\ref{aeps}) gives the rate of change of 
their real parts:
\begin{equation}
\frac{\partial u}{\partial {\frak h}_2} = \pm 
   \frac{\pi}{2} \frac{\gamma}{\sqrt{\epsilon}} \, {\rm sech\/} x
   + O(\gamma^2).
   \label{uh}
\end{equation}
As $\epsilon \to 0$, the rate of change becomes 
infinitely large.
 Away from the neighbourhood of the
turning point, in the region of the applicability of the regular expansion
(\ref{reg}),
the rate of the transformation 
of the solitons $\psi_{\pm}$
is given by   
$$
 \frac{d \alpha}{d {\frak h}} =
 \frac{2}{\pi {\frak h}} \frac{1}{\sqrt{{\frak h}^2 - (2 / \pi)^2}}.
$$
Similarly to Eq.(\ref{uh}),  this shows that as ${\frak h} \to 2/ \pi$,
the two solitons transform increasingly fast.
 
 Thus if we want to use the two real solitons (\ref{g=0}) 
 as approximations for their respective
  ($\gamma \ne 0$)-counterparts, we
 should keep in mind that this approximation is valid only far
 away from the turning point $\gamma_{\rm thr}$. Since for small $h$ 
 the turning point is close to the $h$-axis (i.e. $\gamma_{\rm thr}
 \sim (\pi/2) h$ is also small), the validity of the approximation
 will be restricted to {\it very\/} small $\gamma$: $\gamma/h \ll \pi/2$.

\noindent {\bf 2.} 
Finally, we briefly summarize the main points of this work.

{\bf (a.)}
The lower boundary of the existence domain of the two solitons
is given by the following asymptotic expression (as $\gamma \to 0$):
\begin{equation}
h_{\rm thr} = \frac{2}{\pi} \gamma + (1.667 \times 10^{-3}) \gamma^3
+... .
\end{equation}

{\bf (b.)} 
For $h$ away from the above threshold, more precisely for
$h-(2/ \pi)\gamma \gg 0.002 \gamma^3$, 
 the solitons are given by the asymptotic expansion Eq.(\ref{reg})
 where $v_1$ and $u_1$ are given by explicit expressions
 (\ref{v_1})-(\ref{u1}) with $A^{(0)} = -2.4378 \times 10^{-1}$.

{\bf (c.)} 
For $h$ close to the turning point, $h=(2/\pi)\gamma +{\frak h}_2 \gamma^3$
with ${\frak h}_2 \sim 10^{-3}$
the second, $\gamma^2$-order of the regular expansion (\ref{reg})
becomes greater than the first order, and the expansion breaks down.
In this case the two solitons are given by the singular expansion
(\ref{right}) with  $A^{(\pm)}$ as in Eq.(\ref{Apm}) 
and ${\frak h}_2^{(0)}=1.667
\times 10^{-3}$.

\section{Acknowledgments}
We thank Dmitry Pelinovsky for  reading the manuscript and making
a useful suggestion (which gave rise to the unified formalism
of Sec.\ref{unified}).
  Michael Kollmann's and Mikhail Bogdan's
remarks and Nora Alexeeva's computational
assistance  are also highly appreciated.
I.B. is grateful to George Tsironis, Theo Tomaras and Nikos Flytzanis
for their hospitality at the University of Crete.
 This research was supported by the FRD of South
Africa and  the URC of the University of
Cape Town. 
I.B. was also supported by the FORTH of Greece and
E.Z. was supported by an RFFR grant $\#$RFFR 97-01-01040.

\end{multicols} 

\end{document}